\documentclass[namedreferences]{solarphysics}
\usepackage[hyperref,optionalrh,solaromanenum]{spr-sola-addons} % For Solar Physics 
\usepackage{graphicx}                    % For eps figures, newer & more powerfull
\usepackage{amssymb}                    % useful mathematical symbols
\usepackage{color}                       % For color text: \color command
\usepackage{breakurl}                         % For breaking URLs easily trough lines
\newcommand{\lya}{Ly$\alpha$}

\begin{document}

\begin{article}
\begin{opening}

\title{Solar Irradiance Variability Due To Solar Flares Observed in Lyman-alpha Emission}
%\title{Solar Spectral Irradiance Variability Due To Solar Flares Observed in Lyman-alpha Emission}
%\title{The Contribution of Solar Flares Observed in Lyman-alpha Emission to the Solar Spectral Irradiance}
%\title{Superposed-Epoch Analysis of Solar Flares Observed in Lyman-alpha' Emission}

%%%%%%%%%%%%%%%%%%%%%%%%%%%%%%%%%%%%%%%%%%%%%%%%%%%
%% Authors Names
%
\author[addressref={aff1},email={r.milligan@qub.ac.uk}]{\inits{R.O.}\fnm{Ryan~O.}~\lnm{Milligan}\orcid{0000-0001-5031-1892}}

%%%%%%%%%%%%%%%%%%%%%%%%%%%%%%%%%%%%%%%%%%%%%%%%%%%
%% Runningheads
%
\runningauthor{Milligan}
\runningtitle{Solar Irradiance Variability Due To Solar Flares Observed in Lyman-alpha Emission}

%%%%%%%%%%%%%%%%%%%%%%%%%%%%%%%%%%%%%%%%%%%%%%%%%%%
%% Affilations 
%% id sholud be the same with \author addressref value.
\address[id={aff1}]{Astrophysics Research Centre, School of Mathematics and Physics, Queen's University Belfast, University Road, Belfast, BT7 1NN, Northern Ireland}
%\address[id={aff2}]{Astrophysics Research Centre, School of Mathematics and Physics, Queen's University Belfast, University Road, Belfast, BT7 1NN, Northern Ireland}
%\address[id={aff3}]{Astrophysics Research Centre, School of Mathematics and Physics, Queen's University Belfast, University Road, Belfast, BT7 1NN, Northern Ireland}

%%%%%%%%%%%%%%%%%%%%%%%%%%%%%%%%%%%%%%%%%%%%%%%%%%%
%%% Abstract 
\begin{abstract}
As the Lyman-alpha (\lya) line of neutral hydrogen is the brightest emission line in the solar spectrum, detecting increases in irradiance due to solar flares at this wavelength can be challenging due to the very high background. Previous studies that have focused on the largest flares have shown that even these extreme cases generate enhancements in \lya\ of only a few percent above the background. In this study, a superposed-epoch analysis was performed on $\sim$8500 flares greater than B1 class to determine the contribution that flares make to changes in the solar EUV irradiance. Using the peak of the 1--8\AA\ X-ray emission as a fiducial time, the corresponding time series of 3123 B- and 4972 C-class flares observed in \lya\ emission by the EUV Sensor on GOES-15 were averaged to reduce background fluctuations and improve the flare signal. The summation of these weaker events showed that they produced a 0.1--0.3\% enhancement to the solar \lya\ irradiance. For comparison on average, the same technique was applied to 453 M- and 31 X-class flares, which resulted in a 1--4\% increase in \lya\ emission. Flares were also averaged with respect to their heliographic angle to investigate any potential center-to-limb variation. For each GOES class, the relative enhancement in \lya\ at the flare peak was found to diminish for flares that occurred closer to the solar limb due to the opacity of the line, and/or foreshortening of the footpoints. One modest event included in the study, a C6.6 flare, exhibited an unusually high increase in \lya\ of 7\% that may have been attributed to a failed filament eruption. Increases of this magnitude have hitherto only been associated with a small number of X-class flares. 
\end{abstract}

%%%%%%%%%%%%%%%%%%%%%%%%%%%%%%%%%%%%%%%%%%%%%%%%%%%
%% Keywords
%
%\keywords{}

\end{opening}
%-------------------------------------------------

%%%%%%%%%%%%%%%%%%%%%%%%%%%%%%%%%%%%%%%%%%%%%%%%%%%
%% Sections
%
\section{Introduction}\label{sec:intro}
The Lyman-alpha (\lya) line of neutral hydrogen at 1216\AA\ is the brightest emission line in the solar spectrum. It results from the 2p--1s transition, and in the quiet-Sun, its formation height ranges from the mid-chromosphere ($\sim$6,000~K; line wings) to the base of the transition region ($\sim$40,000~K; line core). During solar flares, nonthermal electrons deposit their energy at these layers, generating localised heating and ionisation at the flare footpoints. Precisely how the \lya\ line responds to this injection of energy remains unknown due to the lack of spectrally-resolved measurements of this fundamental line during flares. One exception is a serendipitous observation of the 28 October 2003 X17 flare by \cite{wood04}. They found that the line core increased by around 20\% while the line wings were enhanced by around factor of two, with the blue wing responding more than the red wing (see also \citealt{canf80}, \citealt{brek96}, and \citealt{mill16}). However, even in this case the data were recorded with a scanning spectrograph and so different parts of the line profile may have been sampled at different stages of the flare's impulsive phase.

\lya\ is believed to be a significant radiator of flare energy, accounting for several percent of the total radiated energy in a single emission line \citep{daco09,mill14}. It has also recently been shown that flare-induced acoustic oscillations can be detected in disk-integrated \lya\ observations, suggesting a potential diagnostic for the dissipation of mechanical energy as well \citep{mill17,mill20}. The study of flares in \lya\ is also crucial for the field of space weather as photons at this wavelength can induce chemical and dynamic changes in planetary atmospheres \citep{chub57,lean85,wood95}. During quiescent solar conditions, \lya\ photons get absorbed in the D-layer of Earth's ionosphere, along with solar X-rays, although \cite{raul13} were unable to detect any appreciable D-layer effects during seven small-to-moderate flares observed in \lya; these flares were found to produce \lya\ enhancements of $<1$\%. However, during an X-class flare that exhibited a $\sim$30\% \lya\ enhancement, \cite{mill20} noted that the impulsive increase in \lya\ emission correlated with induced currents in the E-layer due to the ionization of nitric oxide as determined from magnetometer data. The corresponding D-layer response occurred several minutes later in line with the more gradually varying soft X-rays (SXR). Further study of flares in \lya\ emission and the corresponding response of the terrestrial atmosphere is required to establish if was an isolated case, or a more common occurrence.

The few reports of solar flare observations in \lya\ that previously existed were mostly from broadband, disk-integrated irradiance measurements, and often focused on small numbers of events due to the limited duty cycles of instruments capable of observing flares at this wavelength \citep{kret13,raul13,mill14,kret15,mill16,domi18}. However, \cite{mill20} recently carried out a statistical study of 477 M- and X-class flares in broadband, disk-integrated \lya\ emission, and calculated the distribution of contrasts. It was found that most flares (95\%) produced less than a 10\% enhancement above background, with a maximum of $\sim$30\%, although X-class flares that occurred closer to the solar limb were found to produce less of an enhancement. They also measured how  the total energy radiated in \lya\ scaled with energy radiated in X-rays; the excess solar irradiance in \lya\ was found to be up to two orders of magnitude more energetic than that of the corresponding X-rays.

The study of \cite{mill20} focused on larger flares as smaller GOES class events (B- and C-class) did not produce an appreciable response in \lya\ above the intense background. This may be due to instrumental sensitivity, or perhaps \lya\ emission from weaker events is not inherently significant. The aim of this paper is therefore to use a superposed-epoch analysis technique to determine how much of an enhancement these smaller flares would produce, if any, and how much this increase contributes to the solar irradiance at this wavelength. \cite{kret10} used this simple technique to determine the effects of solar flares on the total solar irradiance (TSI; solar flux over all wavelengths incident at the top of Earth's atmosphere). Using the time of peak soft X-ray emission (i.e. the 1--8\AA\ channel from the X-Ray Sensor (XRS) on-board the Geostationary Operational Environmental Satellite (GOES; \citealt{hans96})) as a key time marker for several hundred flares, the corresponding TSI time series around each flare could then be co-added (or averaged) to reduce the background fluctuations (due to acoustic oscillations and granulation), thereby increasing the coherent flare signal. Similarly, \cite{kret11} followed this study by using the same technique on visible, EUV, and SXR irradiance datasets from the Solar and Heliospheric Observatory (SOHO) to show that all flares can be considered to be white light flares, and that the visible component of flare emission (corresponding to a temperature of 9,000~K) can amount to 70\% of the total radiated energy. Section~\ref{sec:data_anal} of this paper describes the dataset that was used and how the analysis technique was applied. The results are presented in Section~\ref{sec:results}, while a summary and discussion are given in Section~\ref{sec:conc}.

\section{Data Analysis}\label{sec:data_anal}
The Extreme Ultra-Violet Sensor (EUVS; \citealt{vier07}) onboard the GOES-N series of satellites (GOES-13, GOES-14, and GOES-15) comprises five channels, A, B, C, D, and E, covering the 50--170\AA, 240--340\AA, 200--620\AA, 200--800\AA, and 1180--1250\AA\ wavelength ranges, respectively, with the E-channel centered on the \lya\ line at 1216\AA. Each broadband channel samples the solar EUV irradiance at 10.24~s cadence, compared to the more familiar GOES-N/XRS which samples the 0.5--4\AA\ and 1--8\AA\ wavelength ranges at 2~s cadence. While the availability of EUVS data from GOES-13 and GOES-14 has been inconsistent since their launches, GOES-15 has provided continuous coverage since its launch in 2010. However, only data taken up until 6 June 2016 have been made publicly available at the time of writing. EUVS-E data are scaled to a \textit{Whole Heliosphere Interval} quiet-Sun reference spectrum\footnote{\url{http://lasp.colorado.edu/lisird/data/whi_ref_spectra}} \citep{wood09}, providing \lya\ irradiance measurements in physical units of W~m$^{-2}$. Thus the data may not necessarily reflect flare-related time variations of the line profile, generating some systematic uncertainties. To correct for detector degradation, the EUVS-E data are scaled to the daily-averaged data from the Solar–Stellar Irradiance Comparison Experiment (SOLSTICE; \citealt{mccl05}) aboard the Solar Radiation and Climate Experiment (SORCE), and only Version 4 of the EUVS-E data were used in this study\footnote{\url{https://www.ngdc.noaa.gov/stp/satellite/goes/doc/GOES_NOP_EUV_readme.pdf}}.

\begin{figure}
%\begin{center}
\includegraphics[width=\textwidth]{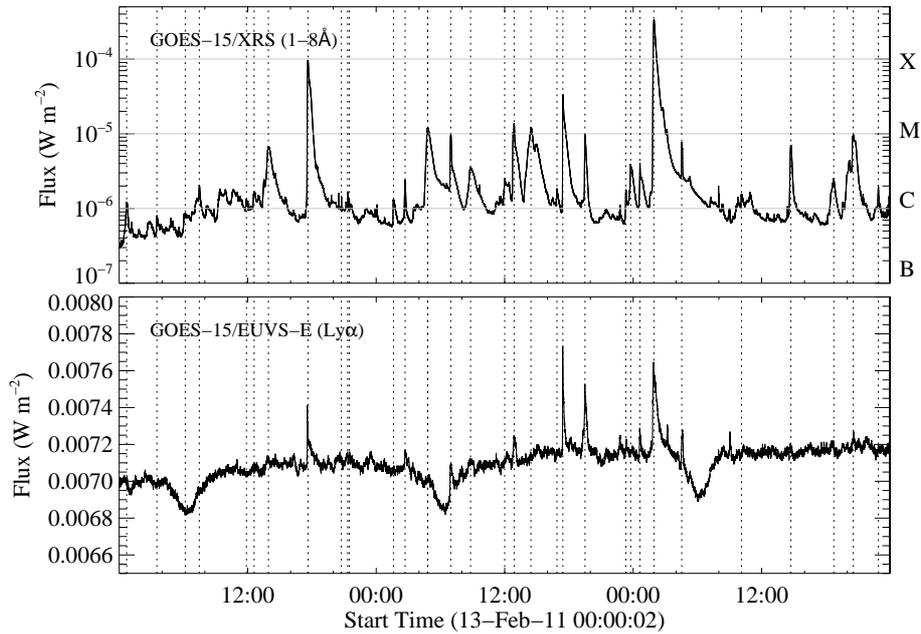}
\caption{72 hours of X-ray (top panel) and \lya\ (bottom panel) irradiance data from GOES-15 during February 2011. The vertical dotted lines in each panel denote the peak times of all events listed in the NOAA flare catalog. The daily dips in the \lya\ data are due to geocoronal absorption by the Earth's outer atmosphere.}
\label{fig:72hr_goes}
%\end{center}
\end{figure}

\begin{figure}
%\begin{center}
\includegraphics[width=\textwidth]{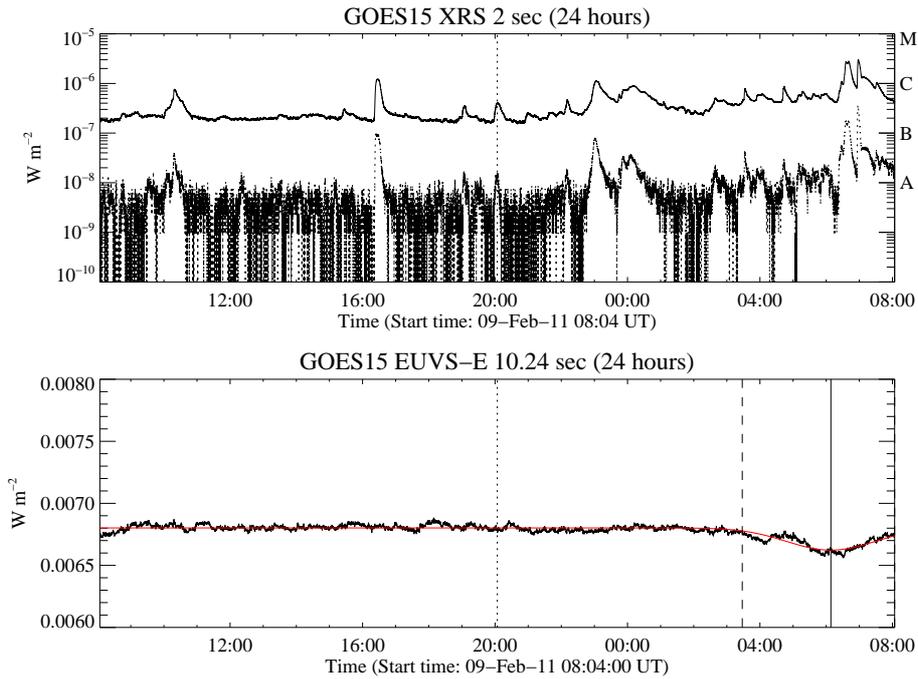}
\caption{Top panel: 24 hours of GOES-15/XRS data at 1--8\AA\ (solid curve) and 0.5--4\AA\ (dotted curve) centred on the peak of the B2.9 flare that occurred on 9 February 2011 (SOL2011-02-09) beginning at 19:57~UT. The vertical dotted line denotes the peak of the B2.9 flare as listed in the NOAA event list. Bottom panel: The GOES-15/EUVS-E (\lya) data over the same 24 hour period. The red line marks the background value over this period, taken as a constant equal to the modal value plus a Gaussian to account for geocoronal absorption, seen as a dip in the data around 06:00~UT on 10 February 2011. The vertical solid line marks the minimum of the dip, while the vertical dashed line marks the 2$\sigma$ width.}
\label{fig:bflare_ex}
%\end{center}
\end{figure}

Figure~\ref{fig:72hr_goes} shows a 72-hour period from February 2011, during which, flares of a range of X-ray magnitudes occurred (top panel; the vertical dotted lines in both panels denote the peak times of each X-ray event). However, over the same period, only the largest of these events produced a discernible response in \lya\ emission (bottom panel). Also visible in the bottom panel are the daily dips in detected \lya\ emission due to attenuation by the Earth's geocorona - which is opaque to \lya\ photons \citep{meie70,bali19} - for a few hours during each GOES orbit. For every event considered in this study, a 24 hour period of \lya\ emission centred on the peak X-ray time of the flare was fitted with a constant, equal to the modal value, plus an inverted Gaussian (red curve in the bottom panel of Figure~\ref{fig:bflare_ex}) to account for the geocoronal dip given that the amount of attenuation varies throughout the year. As the effect of the geocorona on \lya\ flare irradiance is likely to be nonlinear, any events that occurred within $\pm2\sigma$ (vertical dashed line) of the minimum of each daily dip (vertical solid line), were omitted from this study. 

The start, peak, and end times of every X-ray flare are listed in the GOES event list\footnote{\url{https://www.ngdc.noaa.gov/stp/satellite/goes/index.html}} hosted by the National Oceanic and Atmospheric Administration/Space Weather Prediction Center (NOAA/SWPC). However this list does not always include a flare's location, whereas the Heliophysics Event Knowlegebase (HEK)\footnote{\url{https://www.lmsal.com/hek/}} does, although it is missing several months of data throughout Solar Cycle 24 \citep{mill18}. The superposed-epoch technique employed in this study uses the peak time of the of all X-ray flares above B1 from the HEK database as a fiducial time after removing any events affected by geocoronal absorption, eclipse periods, or corrupted data. This resulted in 3123 B-class and 4972 C-class flares\footnote{Note that there were more C-class flares than B-class flares due to the high X-ray background around the time of solar maximum.}. While the \lya\ component of M- and X-class flares can often be detected above the solar background, the averaged profiles of 453 M- and 31 X-class flares were also included in this analysis. This then allows the average flux (in W~m$^{-2}$) and contrast (relative percentage increase) to be measured for each GOES classification (Section~\ref{sec:results}). Furthermore, given the abundance of events observed by GOES-15/EUVS-E over Solar Cycle 24, and with knowledge of their locations from the HEK, the average \lya\ profile can be determined as a function of heliographic angle in order to establish any potential center-to-limb variation (CLV; Section~\ref{sec:clv}).

\begin{figure}[!h]
%\begin{center}
\includegraphics[width=\textwidth]{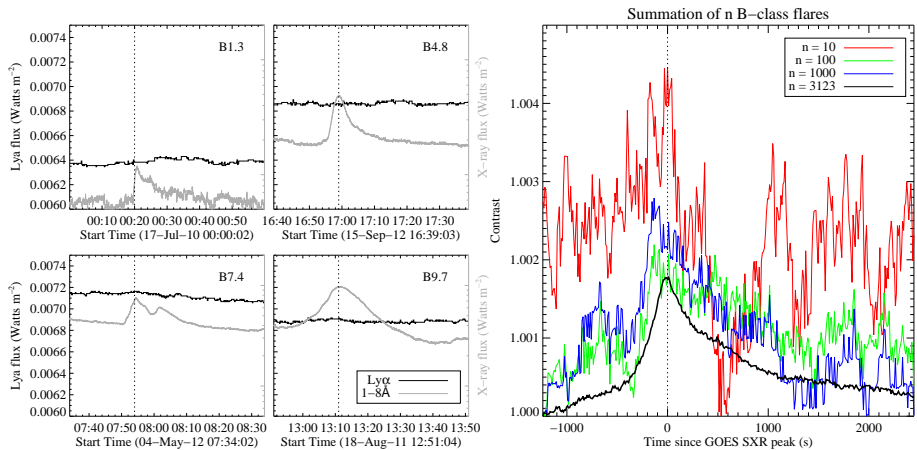}
\caption{Left hand panels: Time profiles for four B-class flares in \lya\ (black) and X-rays (grey). The scaling is the same on each panel. The vertical dotted line on each panel marks the X-ray peak as catalogued in the NOAA flare list. The right hand panel illustrates how the \lya\ flare signal becomes more apparent when adding together increasing numbers of events. The red curve is the average of 10 random B-class flares; the green curve is a result of 100 events; the blue curve, 1000; and finally the black curve is an average of all 3123 B-class flares from this study.}
\label{fig:nbflares}
%\end{center}
\end{figure}

A one-hour time range of the \lya\ time series, from 20 minutes prior to the X-ray peak to 40 minutes after, were averaged over for all flares of a given GOES class. Four B-class flares are shown in the left hand panels of Figure~\ref{fig:nbflares} as an example. For each individual event, there does not appear to be any discernible \lya\ enhancement due to the flare, but by averaging over increasing numbers of \lya\ time series of equal length, the signal-to-noise ratio is vastly improved as shown in the right hand panel. The red, green, and blue profiles represent the summation of 10, 100, and 1000 randomly chosen B-class flares, respectively. The smoothly-varying solid black lightcurve is the result of averaging all 3123 B-class flares considered. This illustrates how the signal-to-noise ratio is increased by increasing the number of events added together. 

\section{Results}\label{sec:results}

\begin{figure}[!t]
%\begin{center}
\includegraphics[width=0.48\textwidth]{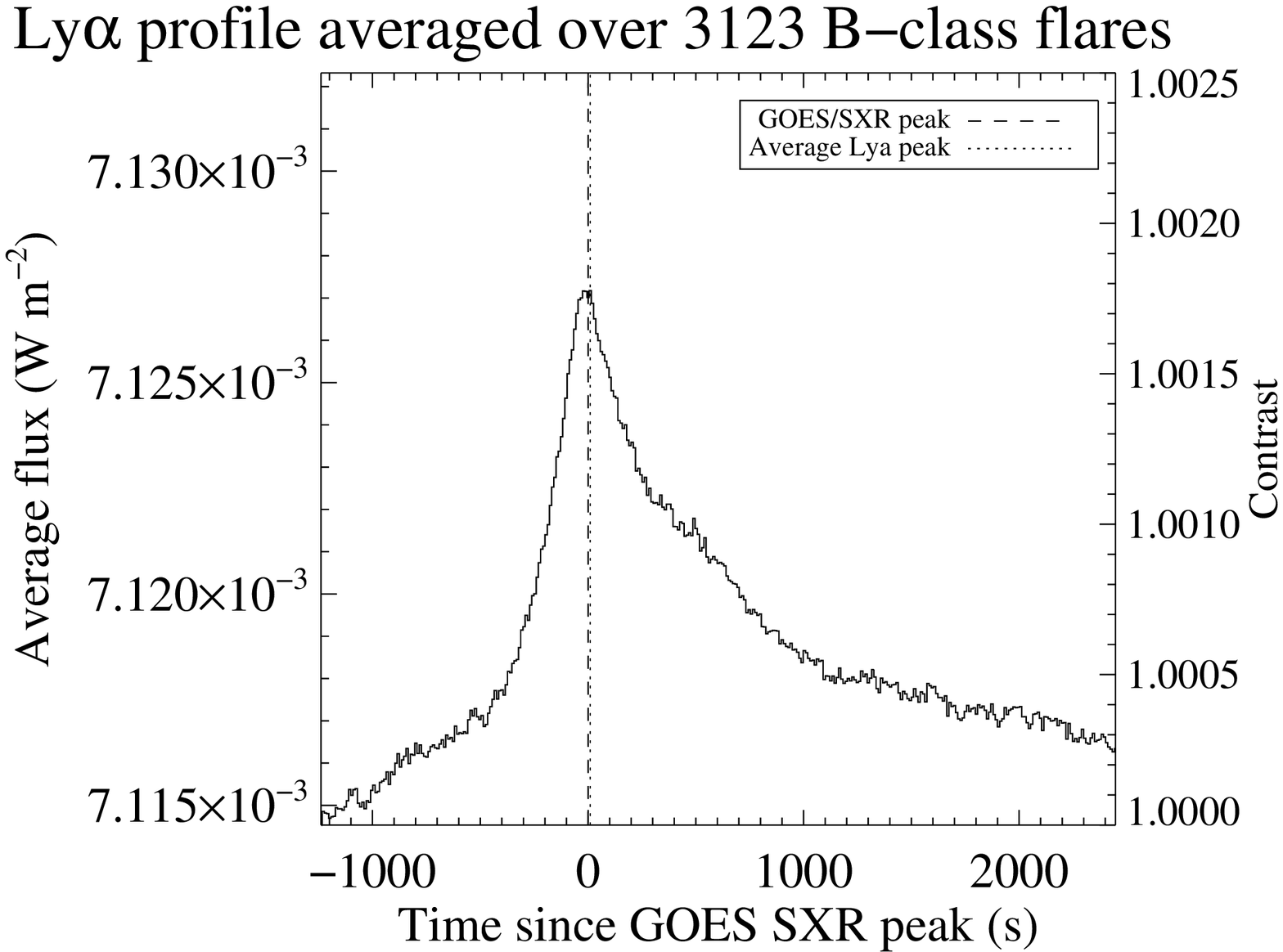}
\includegraphics[width=0.48\textwidth]{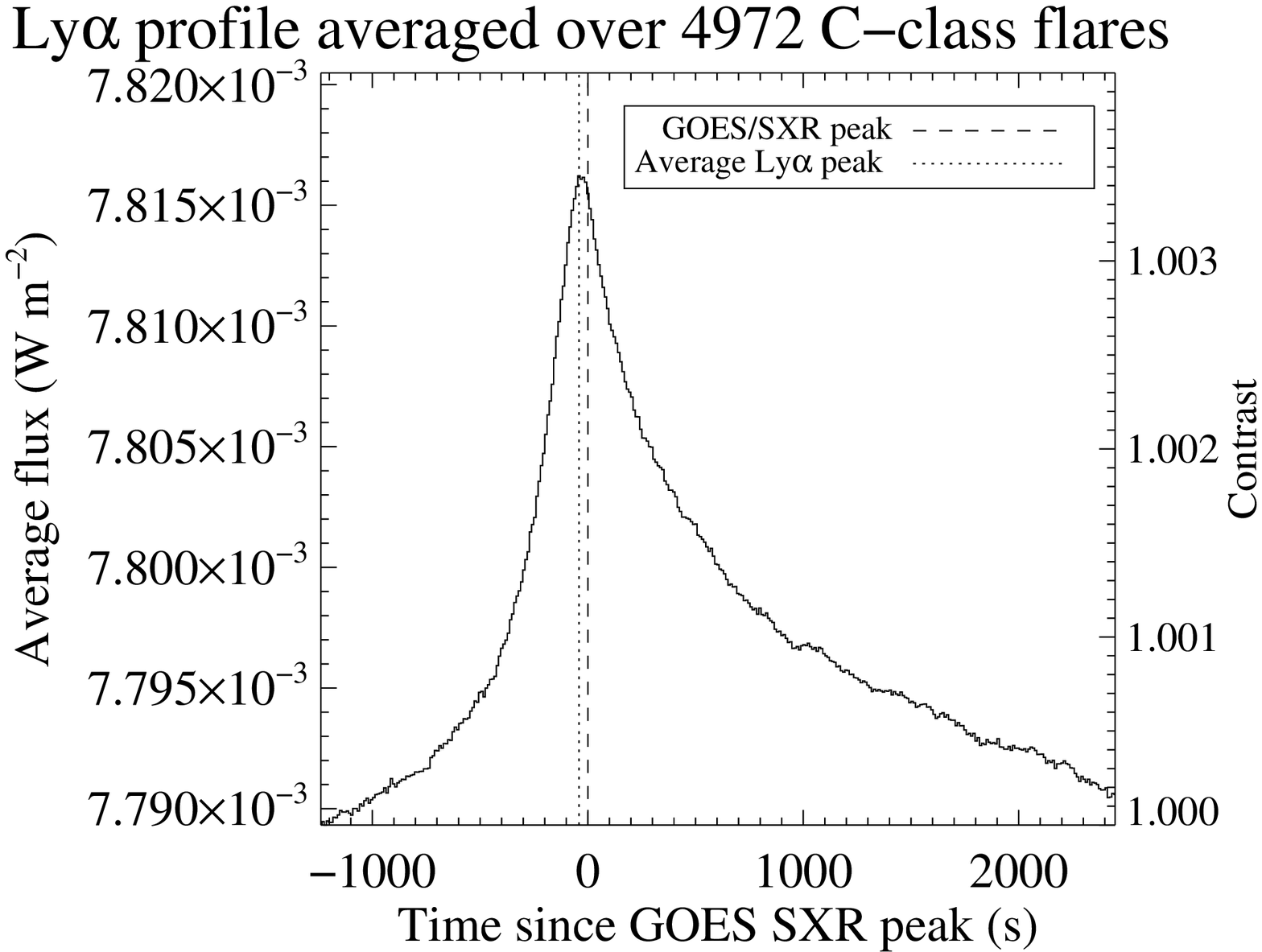}
\includegraphics[width=0.48\textwidth]{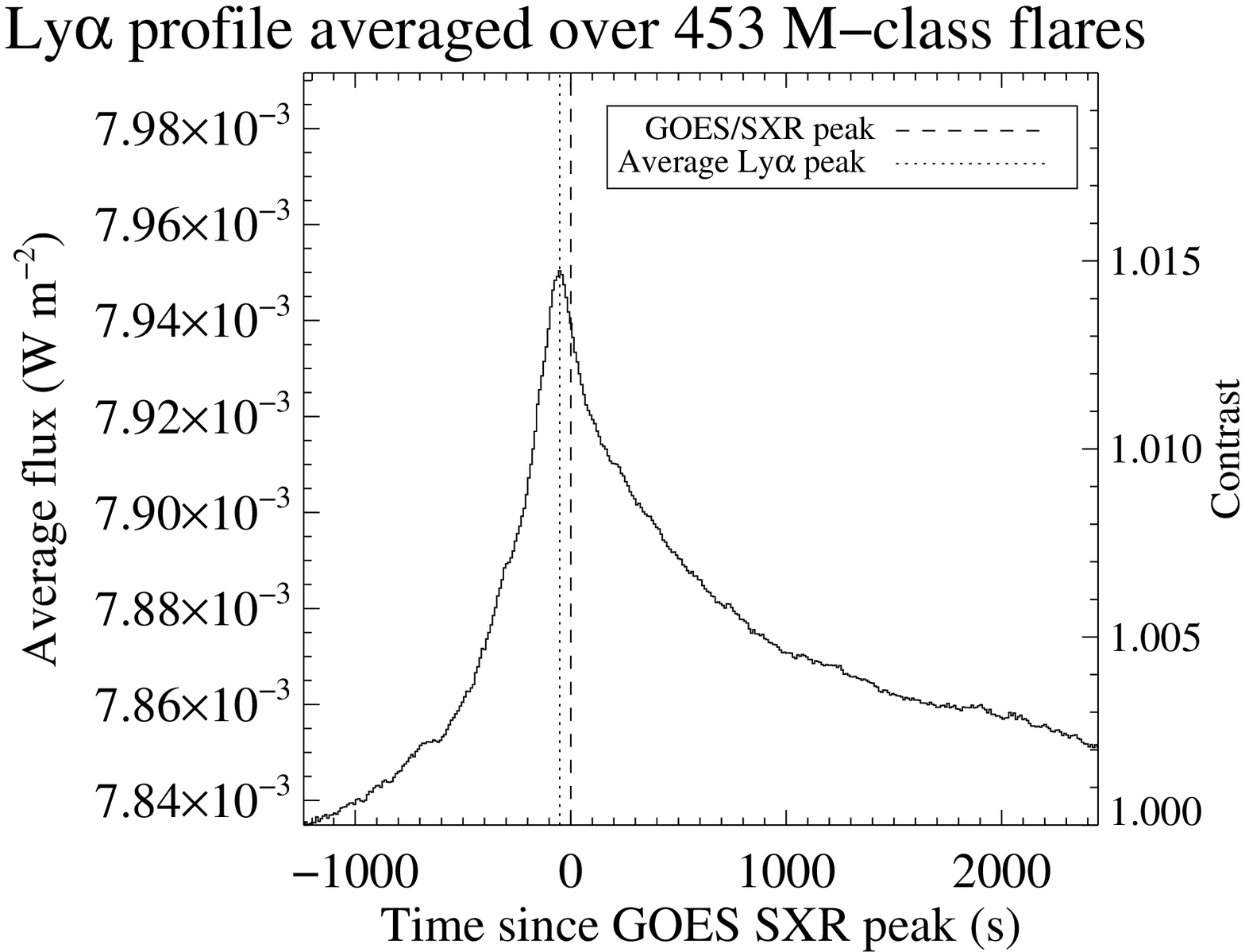}
\includegraphics[width=0.48\textwidth]{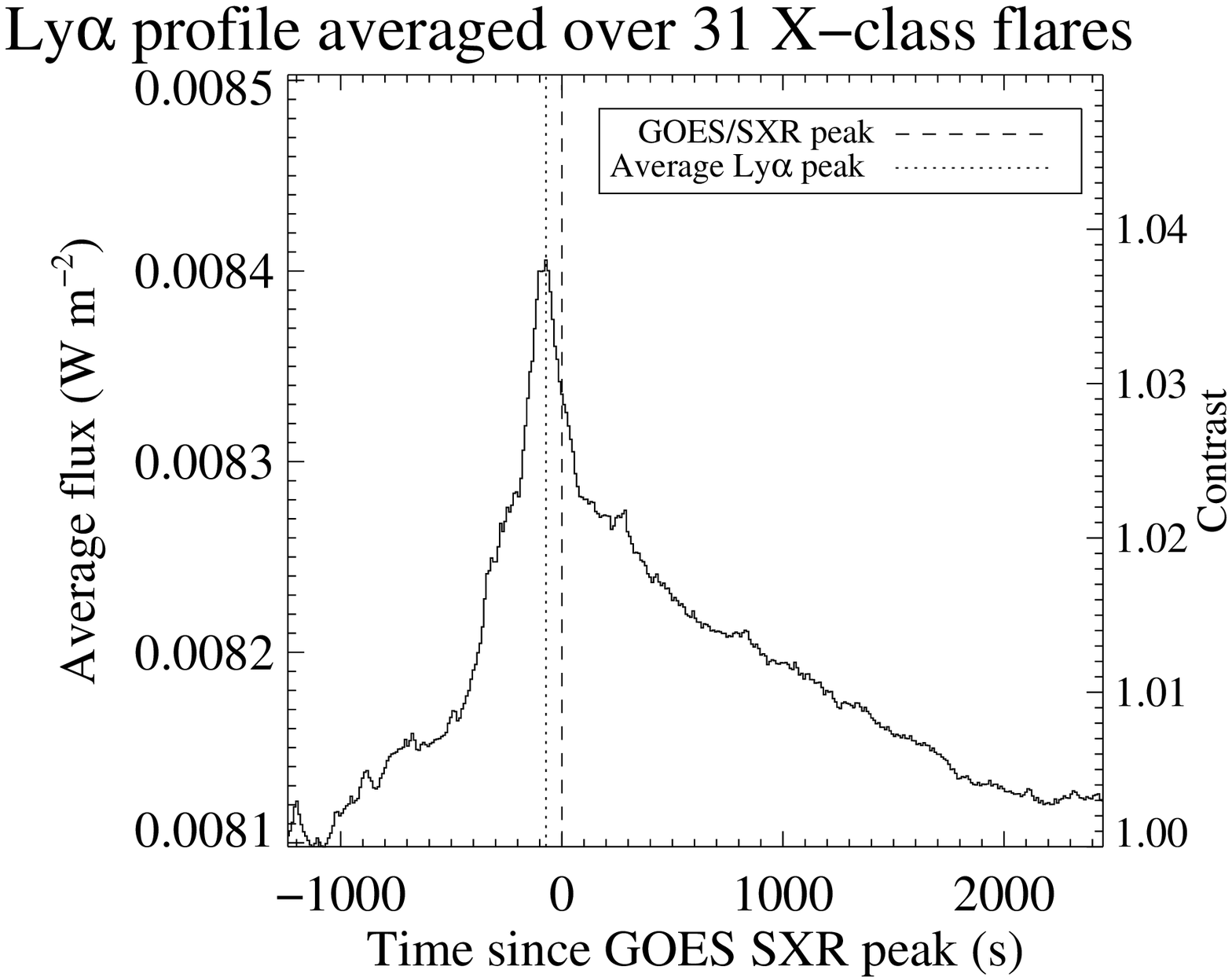}
\caption{Plots of the average time profiles for all B-, C-, M-, and X-class flares included in this study. The vertical dashed line in each panel marks the time of the SXR peaks of all flares, while the vertical dotted line marks the peak of the \lya\ emission. The contrast axis denotes the percentage increase in each panel.}
\label{fig:spe_flares}
%\end{center}
\end{figure}

The four panels in Figure~\ref{fig:spe_flares} show the average \lya\ time profiles for B-, C-, M-, and X-class flares. All profiles show a clear and distinct peak in \lya\ emission that was not readily visible for individual events, particularly weaker GOES classes. The average B-class flare produced an enhancement of 0.18\% above background, while C-class flares showed a 0.35\% increase. These values are in agreement with those reported by \cite{raul13} for several C-class flares observed by PROBA-2/LYRA. Most individual M- and X-class flares can be readily observed in \lya, as shown by \cite{mill20}, but are included here for completeness. M-class flares were measured to increase the solar irradiance by 1.5\% on average, while X-classes exhibited a 3.8\% increase in \lya\ above background. Although the flare-related changes in \lya\ irradiance quoted here are quite small, these changes can still correspond to a significant amount of radiated energy. This confirms that weaker flares do have associated responses in \lya\ emission that are ordinarily obscured by the intense solar background. This will have significant implications for more sensitive \lya\ instruments that aim to observe solar flares in the future.

The peak timings of the averaged profiles are also shown in Figure~\ref{fig:spe_flares} as vertical dotted lines. For B-class flares, the average peak time coincides with the peak of the SXR emission for the summed events to within one GOES-15/EUVS-E time bin (10.24~s). For C-class flares, \lya\ peaks slightly earlier than the X-ray peak (40.96~s; four time bins), while M- and X-class flares peak even earlier still (51.20~s and 71.68~s, respectively). This somehow implies that \lya\ emission from B-class flares may be thermally generated, while C-, M-, and X-class flares exhibit more nonthermal \lya\ emission in accordance with the Neupert Effect \citep{neup68}. Although it is tempting to assume that this systematic change in peak timing is physically significant, great care must be taken in drawing conclusions from summed epoch analyses as individual events can sometimes skew the resulting profile.

\subsection{Centre-to-Limb Variation}\label{sec:clv}

\begin{figure}[!b]
\includegraphics[width=\textwidth]{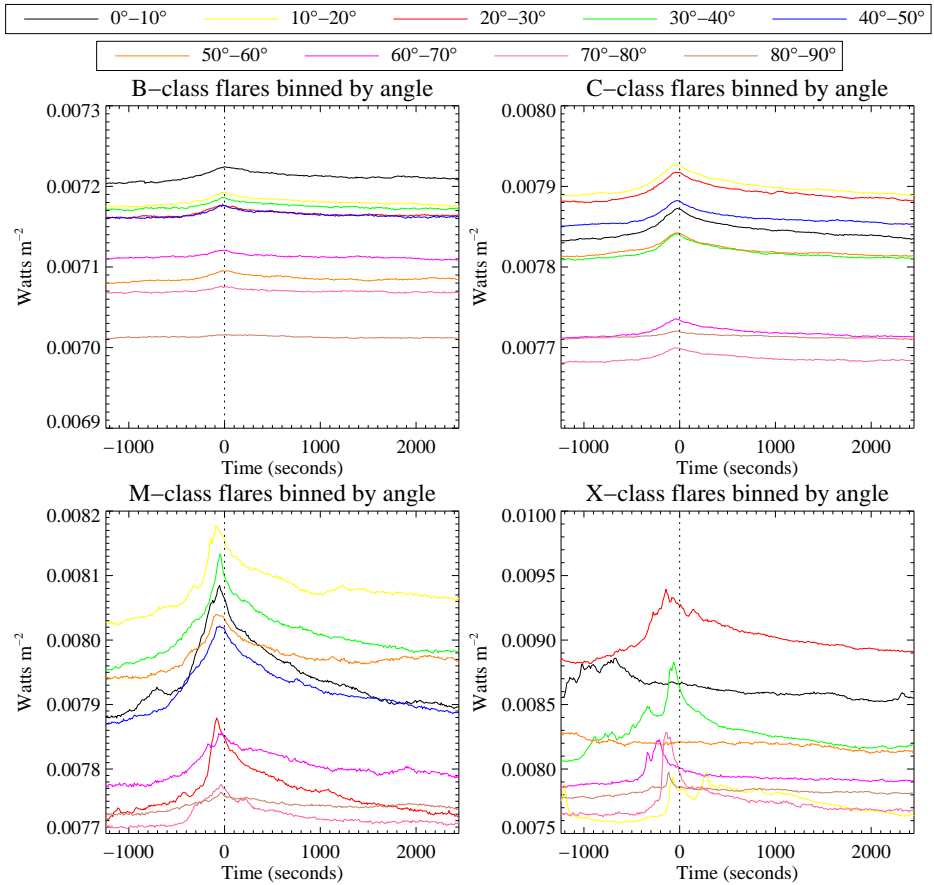}
\caption{Plot of the \lya\ time profiles for flares binned by heliographic angle in increments of 10 degrees for B-, C-, M-, and X-class flares. The vertical dashed line in each panel marks the time of the SXR peaks of all flares.}
\label{fig:mplot_clv_cont}
\end{figure}

As \lya\ is optically thick \citep{wood00}, flares that occur closer to the solar limb are assumed to produce less of a response in \lya\ than if they had occurred at disk centre due to scattering by the increasing column mass along the line-of-sight. \cite{mill20} showed that this was true for 31 X-class flares by normalising their \lya\ flux to their X-ray flux (which is optically thin) and showing that this ratio decreases with increasing heliocentric angle from disk centre. This center-to-limb variation was also confirmed through a stereoscopic observation of an X-class limb event as viewed by GOES-15 from Earth that was observed simultaneously by the EUV Monitor on-board the Mars Atmospheric and Volatile Evolution satellite (MAVEN/EUM; \citealt{epar15}), when Earth and Mars were around 90 degrees apart. After correcting for the Earth-Mars distance and light travel time, the flux detected by MAVEN (which observed the flare close to disk center) was greater than that detected by GOES.

\begin{figure}[!b]
\begin{center}
\includegraphics[width=0.75\textwidth]{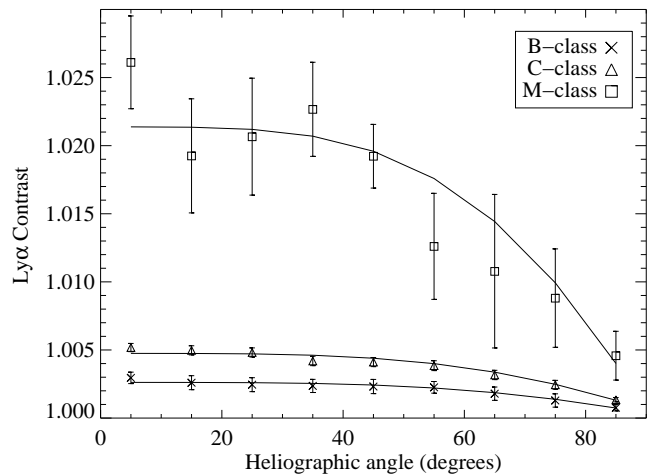}
\caption{Plot of the \lya\ contrast for each of the averaged lightcurves in Figure~\ref{fig:mplot_clv_cont} (excluding X-classes) as a function of heliographic angle. The curves fitted to each GOES class are given by Equation~\ref{eqn:clv}.}
\label{fig:flare_clv_cont}
\end{center}
\end{figure}

Figure~\ref{fig:mplot_clv_cont} shows averaged flare time profiles for the four GOES classes in increments of 10 degrees from 0--90$^{\circ}$. In most cases, flares that occurred closer to disk center had a higher absolute flux value, and a more pronounced peak relative to the background around the time of the GOES SXR peak. This trend breaks down for X-class flares due to the diminishing number of events. To illustrate this point, Figure~\ref{fig:flare_clv_cont} shows the contrast value (peak flux divided by minimum flux) for each heliographic angle bin for B-, C-, and M-class flares. Error bars were taken as the 1$\sigma$ value for each curve in Figure~\ref{fig:mplot_clv_cont} divided by the square root of the number events in each bin \citep{kret10}. These data points were fitted with the commonly used CLV function: 

\begin{equation}  
R = R_C\left(k+2(1-k)\left(\mu-\frac{\mu^2}{2}\right)\right),
\label{eqn:clv}
\end{equation}

\noindent
as derived by \cite{brek94} and also employed by \cite{wood06} and \cite{mill20}. $R$ is the intensity at a given angle, $R_C$ is the intensity ratio at disk center, $k$ is the limb variation relative to center, and $\mu=cos(\theta)$. The values of $R_C$ for B-, C-, and M-class flares are 1.002, 1.004, and 1.021, respectively, while the corresponding $k$ values are 0.99, 0.99, and 0.97. In agreement with \cite{mill20}, all flare classifications showed a smaller relative enhancement in \lya\ emission for flares that occurred closer to the solar limb compared to those that occurred closer to disk center. While it would be more physically meaningful to normalize each curve to their respective GOES class \citep{wood06}, the summed nature of the superposed-epoch analysis coupled with the difficulty associated with subtracting the solar background for smaller X-ray events, make this much more of a challenge, and is likely to return similar trends, albeit with much lower $k$-values. Note that \cite{kret10} and \cite{kret11} did not include flares that occurred beyond 60$^{\circ}$ of disk center in their superposed-epoch analysis on flares in the TSI, visible, EUV, and SXR emission as they assumed that limb flares would not produce a measurable response at these wavelengths in the chromosphere.

\subsection{An Unusual C-class Flare}\label{sec:cclass}

\begin{figure}[!b]
%\begin{center}
\includegraphics[width=\textwidth]{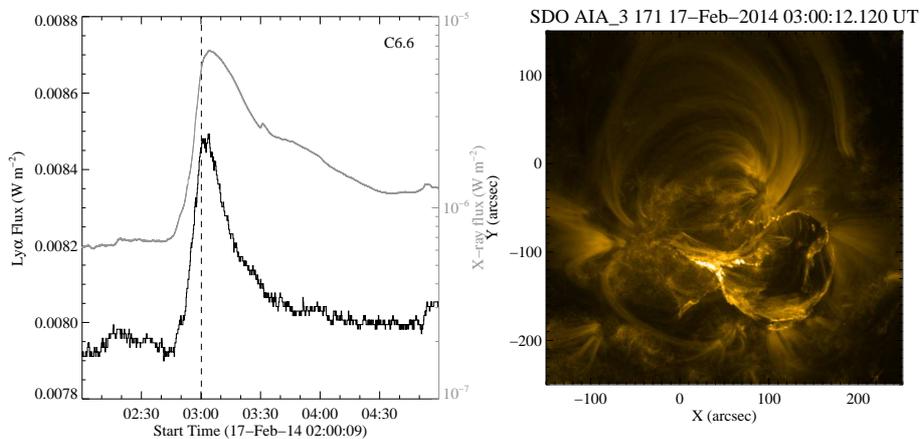}
\caption{Left: Plot of \lya\ (black) and X-ray (grey) time profiles for a C6.6 flare that occurred on 17 February 2014. The vertical dashed line denotes the time of the SDO/AIA image in the right hand panel. Right: SDO/AIA 171\AA\ image taken around the time of peak \lya\ emission showing the possible failed filament eruption that may be responsible for the unusually high \lya\ flux.}
\label{fig:c6_flare}
%\end{center}
\end{figure}

While the vast majority of B- and C-class flares do not produce a \lya\ signature above the bright solar background, one event in this study - a C6.6 flare that occurred on 17 February 2014 (SOL2014-02-17)- displayed a remarkable 7\% enhancement. Such a degree of contrast has hitherto only been associated with a few X-class flares \citep{wood04,mill20}. Converting the flare excess in \lya\ into units of energy for this event yielded a value of $\sim$10$^{30}$~erg, which is comparable to that calculated for the 15 February 2011 X2.2 by \cite{mill14} using \lya\ data from the EUV Variability Experiment (EVE; \citealt{wood12}) on-board the Solar Dynamics Observatory (SDO; \citealt{pesn12}). A plot of the \lya\ and X-ray lightcurves for the event is shown in Figure~\ref{fig:c6_flare}, along with a 171\AA\ image from the Atmospheric Imaging Assembly (AIA; \citealt{leme12}), also on SDO, taken around the time of peak emission (vertical dashed line in left hand panel). Ordinarily, flare-related increases in \lya\ (or indeed any inherent chromospheric emission) is primarily detected during the impulsive phase as the chromosphere is heated by precipitating particles from the corona, or late in the decay phase as hot flare plasma cools through instrumental passbands. Large increases in \lya\ would therefore usually be attributed to a particularly energetic beam of electrons, and/or an unusually large footpoint area. Unfortunately, hard X-ray data from the Ramaty High Energy Solar Spectroscopic Imager (\citealt{lin02}) or Fermi \citep{meeg09} were not available for this event, and so the parameters of the electron distribution that may have been responsible remains unknown. However, SDO/AIA data appear to show a failed filament eruption in all EUV channels (only 171\AA\ is shown in Figure~\ref{fig:c6_flare}). No associated coronal mass ejection could be seen in coronagraph images around the time and location of the flare. The \lya\ emission from this event appears to correlate well with the main X-ray phase suggesting that the it could have been emanating from the corona as heating took place within the large volume of the filament. A similar conclusion was reached by \cite{daco09} who reported co-spatial \lya\ (from the Transition Region and Coronal Explorer; \citealt{hand99}) and X-ray emission during a filament eruption. 

\section{Conclusions}\label{sec:conc}
This paper presents a study of 8579 solar flares observed in the \lya\ line of neutral hydrogen. As weaker flares do not readily produce a \lya\ signature above the intense solar background at this wavelength, a superposed-epoch analysis was carried out on 3123 B-class flares and 4972 C-class flares. On average, these flares were found to exhibit a 0.1--0.3\% enhancement. For completion and comparison, 453 M- and 31 X-flares were also co-added, and average increases of 1--4\% were measured. For each GOES classification, flares were also summed according to their heliographic location revealing that in all cases, flares closer to disk centre displayed a greater enhancement on average that those that occurred closer to the solar limb due to the opacity of \lya, or possible foreshortening of the flare ribbons. A C-class flare that exhibited an abnormally high contrast of 7\% (equating to 10$^{30}$~erg) was also presented.

While GOES/EUVS was designed as an EUV irradiance monitor, the results presented here show just how stable the data are and how they can be utilized for detailed scientific analysis, in particular for large scale statistical studies. And although the sensitivity of GOES-15/EUVS-E might be limited to larger flares, the new \lya\ irradiance instruments that are part of the EXIS \citep{epar09,cham09} suite on the newly launched GOES-R series of spacecraft (GOES-16 and GOES-17 were launched in February 2017 and June 2018, respectively, with GOES-18 and GOES-19 to follow) will have a greater dynamic range, as well as providing pseudo line profiles by sampling the \lya\ profile in five wavelength pixels rather than as broadband measurements. Similarly, the imaging capability of the EUV Imager (EUI; \citealt{schu11,roch20}) on-board Solar Orbiter will be able to spatially resolve flares in \lya\ for the first time. The findings presented illustrate that even minor flares can produce small, but perceptible changes in the solar \lya\ irradiance, and should therefore serve as a baseline for the advent of new \lya\ flare observations and advanced numerical simulations that will become available during Solar Cycle 25. 

%% Figure 
%
% \begin{figure} }
% \centerline{\includegraphics[width=0.5\textwidth,clip=]{<fig.eps>}}
% \caption{}%\label{fig:?}
% \end{figure}

%% Table
%
% \begin{table}
% \caption{}%\label{tbl:?}
% \begin{tabular}{}     
% \hline
% \multicolumn{2}{c}{<>}
% <data>
% \hline
% \end{tabular}
% \end{table}

%%%%%%%%%%%%%%%%%%%%%%%%%%%%%%%%%%%%%%%%%%%%%%%%%%%%%%%%%%%%%%%%%%%%%%%%%%%
%% Appendix
%
% \appendix   

%%%%%%%%%%%%%%%%%%%%%%%%%%%%%%%%%%%%%%%%%%%%%%%%%%%%%%%%%%%%%%%%%%%%%%%%%%%
%% Acknowledgements
%
\begin{acks}
The author would like to thank Science and Technologies Facilities Council (UK) for the award of an Ernest Rutherford Fellowship (ST/N004981/1), and Hugh Hudson, Mihalis Mathioudakis, Graham Kerr, and Shaun McLaughlin for feedback on an early draft. The GOES data used in this study are freely and publicly available from the instrument websites (GOES/EUVS: \url{https://satdat.ngdc.noaa.gov/sem/goes/data/euvs/GOES_v4/G15/}; GOES/XRS: \url{https://www.ngdc.noaa.gov/stp/satellite/goes/dataaccess.html}).
\end{acks}

%%% %%%%%%%%%%%%%%%%%%%%%%%%%%%%%%%%%%%%%%%%%%%%%%%%%%%%%%%%%%%
%% Bibliography
%
% Using BibTeX
%
\bibliographystyle{spr-mp-sola}
\bibliography{ms.bib}  
%
% Without BibTeX 
% \begin{thebibliography}{}
% \bibitem[\protect\citeauthoryear{Author}{Year}]{key}
%   <bibliographical entry>
%
% \bibitem[\protect\citeauthoryear{}{}]{}
%   
%  
% \end{thebibliography}

\end{article} 
\end{document}